%% file: FUSION2022_ArxivVersion1.tex
\documentclass[conference,10pt]{IEEEtran}
\usepackage{amsmath}
\usepackage{amssymb}
\usepackage{graphicx}
\usepackage{color}
\usepackage{xcolor}
\usepackage{cite}
\usepackage{bm}
\usepackage{epsfig,psfrag}
\usepackage{tabularx}
\usepackage{acronym}
\usepackage[yyyymmdd,hhmmss]{datetime}
\usepackage{bbm}
\usepackage{notation}
\usepackage{mathrsfs} 
\usepackage{bardiag}
\usepackage{tabularx}
\usepackage{multirow}
\usepackage{colortbl,pgfplotstable}

\usepackage[caption=false,font=footnotesize]{subfig}
\usepackage{pgfplots}
\usepackage{tikzscale}
\newcommand{\exportFigures}{true} 

\newcommand{\tikzfolder}{./compiledPlots/}
\input{pgfplot}

\allowdisplaybreaks
\newcolumntype{L}[1]{>{\raggedright\arraybackslash}p{#1}}
\newcolumntype{C}[1]{>{\centering\arraybackslash}p{#1}}
\newcolumntype{R}[1]{>{\raggedleft\arraybackslash}p{#1}}

\providecommand{\ist}{\hspace*{.3mm}}

\providecommand{\rmv}{\hspace*{-.3mm}}
\providecommand{\iist}{\hspace*{1mm}}
\providecommand{\rrmv}{\hspace*{-1mm}}
\providecommand{\nn}{\nonumber}

\newcommand{\T}{\text{T}}

\newcommand{\vu}[2]{\mbox{$#1\,\text{#2}$}} 

\acrodef{pnt}[PNT]{positioning, navigation and timing}
\acrodef{pa}[PA]{physical anchor}
\acrodef{va}[VA]{virtual anchor}
\acrodef{mva}[MVA]{master virtual anchor}
\acrodef{pmva}[PMVA]{potential \ac{mva}}
\acrodef{slam}[RF-SLAM]{Radio frequency simultaneous localization and mapping}
\acrodef{pmf}[PMF]{probability mass function}
\acrodef{pdf}[PDF]{probability density function}
\acrodef{spa}[SPA]{sum-product algorithm}
\acrodef{mmse}[MMSE]{minimum mean-square error}
\acrodef{pde}[PDE]{partial differential equation}
\acrodef{rf}[RF]{radio frequency}
\acrodef{bpf}[BPF]{bootstrap particle filter}
\acrodef{pf}[FP]{particle flow}
\acrodef{mospa}[MOSPA]{mean optimal subpattern assignment}
\acrodef{rmse}[RMSE]{root mean squared error}

\DeclareMathAlphabet{\mathpzc}{OT1}{pzc}{m}{it}

\newdateformat{monthyeardate}{\monthname[\THEMONTH] \THEDAY, \THEYEAR} 

\definecolor{FGgreen}{RGB}{34,139,34}
\definecolor{FGblue}{RGB}{80,120,255}
\definecolor{FGred}{RGB}{255,110,110}

\begin{document}
\title{\vspace{0mm} Data Fusion for Radio Frequency SLAM\\ with Robust Sampling\vspace{-1mm}}

\author{\IEEEauthorblockN{Erik Leitinger\IEEEauthorrefmark{1}\IEEEauthorrefmark{3}, Bryan Teague\IEEEauthorrefmark{2}, Wenyu Zhang\IEEEauthorrefmark{4}, Mingchao Liang\IEEEauthorrefmark{4},  and Florian Meyer\IEEEauthorrefmark{4} \vspace{2mm}}

\IEEEauthorblockA{\IEEEauthorrefmark{1}TU Graz, Austria and \IEEEauthorrefmark{3}Christian Doppler Laboratory for Location-aware Electronic Systems (erik.leitinger@tugraz.at)\vspace{1mm}} 

\IEEEauthorblockA{\IEEEauthorrefmark{2}MIT Lincoln Laboratory, Lexington, MA, USA (bryan.teague@ll.mit.edu) \vspace{1mm}}  

\IEEEauthorblockA{\IEEEauthorrefmark{4}Scripps Institution of Oceanography and Department of Electrical and Computer Engineering} \\[-4.6mm]
\IEEEauthorblockA{University of California San Diego, La Jolla, CA, USA\vspace*{-3mm} (flmeyer@ucsd.edu) }}

\maketitle

\begin{abstract}
Precise indoor localization remains a challenging problem for a variety of essential applications. A promising approach to address this problem is to exchange radio signals between mobile agents and static \acp{pa} that bounce off flat surfaces in the indoor environment. \ac{slam} methods can be used to jointly estimates the time-varying location of agents as well as the static locations of the flat surfaces. Recent work on \ac{slam} methods has shown that each surface can be efficiently represented by a single \ac{mva}.
The measurement model related to this \ac{mva}-based \ac{slam} method is highly nonlinear. Thus, Bayesian estimation relies on sampling-based techniques. The original \ac{mva}-based \ac{slam} method employs conventional ``bootstrap'' sampling. In challenging scenarios it was observed that the original method might converge to incorrect \ac{mva} positions corresponding to local maxima. 
In this paper, we introduce \ac{mva}-based \ac{slam} with an improved sampling technique that succeeds in the aforementioned challenging scenarios. Our simulation results demonstrate significant performance advantages.
\vspace{.5mm}
\end{abstract}

\acresetall
\section{Introduction}
\label{sec:introduction}
\vspace{0mm}

Maintaining accurate and timely situational awareness in indoor environments is challenging in a variety of applications. This is particularly true when no prior geometric information about the environment is available. In such scenarios, it is desirable to generate a map and localize mobile ``agents'' within that map. 
\ac{slam} is a promising methodology for precise localization in the aforementioned scenarios. Here, radio signals that bounce off flat surfaces are used to estimate the locations of mobile agents and the surfaces themselves.

\ac{slam} methods typically represent a flat surface in the environment using the notion of \acp{va}\cite{WitMeiLeiSheGusTufHanDarMolConWin:J16, GentnerTWC2016, LeiMeyHlaWitTufWin:J19, MenMeyBauWin:J19}. A \ac{va} represents the location of the mirror image of a \ac{pa} on a reflecting surface. The reflected propagation path \textit{agent -- surface -- \ac{pa}} can equivalently be described by a direct path \textit{agent -- \ac{va}}. The goal of \ac{slam} methods is to detect and localize \acp{va} along with the time-varying position of the mobile agent \cite{WitMeiLeiSheGusTufHanDarMolConWin:J16,GentnerTWC2016,LeiMeyHlaWitTufWin:J19,MenMeyBauWin:J19}.  By detecting and localizing the \acp{va}, multiple radio propagation paths can be leveraged for agent localization which increases accuracy and robustness.

\subsection{Background and Contributions}

\ac{slam} follows a traditional feature-based SLAM approach \cite{DurrantWhyte2006, Dissanayake2001, MullaneTR2011}, i.e., the map is represented by static \emph{features}, whose unknown positions are estimated using sequential inference. In particular, most \ac{slam} methods consider the \acp{va} as the features to be mapped \cite{LeiMeyHlaWitTufWin:J19,LeitingerICC2019,MenMeyBauWin:J19,MeyGem:J21}. A complicating factor in \ac{slam} is measurement origin uncertainty, i.e., the unknown association of measurements with features \cite{LeiMeyHlaWitTufWin:J19,LeitingerICC2019,MenMeyBauWin:J19,MeyWil:J21}. State-of-the-art techniques for \ac{slam} formulate and solve a high-dimensional sequential Bayesian estimation problem using factor graphs\cite{LeiMeyHlaWitTufWin:J19, LeitingerICC2019,MenMeyBauWin:J19}. Due to the nonlinear measurement model the resulting \ac{spa} relies on sampling techniques \cite{AruMasGorCla:02,WitMeiLeiSheGusTufHanDarMolConWin:J16,GentnerTWC2016,LeiMeyHlaWitTufWin:J19,MenMeyBauWin:J19}. In traditional \ac{slam}, a reflective surface can take part in multiple propagation paths, each path is represented by a \ac{va}, and each \ac{va} is estimated independently \cite{WitMeiLeiSheGusTufHanDarMolConWin:J16,GentnerTWC2016,LeiMeyHlaWitTufWin:J19,MenMeyBauWin:J19}. This limits timeliness and accuracy of \ac{slam}.

The notion of a unique \ac{mva} has been recently introduced to enable data fusion across propagation paths, i.e., to allow multiple paths to be used for joint estimation of a single reflective surface \cite{LeiMey:Asilomar2020_DataFusion}. This approach has the potential to strongly improve the accuracy and convergence time of \ac{slam} \cite{LeiMey:Asilomar2020_DataFusion}.  The original \ac{mva}-based \ac{slam} method \cite{LeiMey:Asilomar2020_DataFusion} employs conventional ``bootstrap'' sampling. In challenging scenarios with one or two \acp{pa} where only range measurements are available, bootstrap sampling has severe limitations. In particular, due to geometric symmetries, \acp{pdf} of \acp{mva} can be multi-modal during some initial time steps, and the sample representations provided by bootstrap sampling may collapse in a wrong mode. In turn, it was observed that the original method might converge to incorrect \ac{mva} positions that are local maxima.

In this paper, we address this limitation of \ac{mva}-based \ac{slam} by introducing an improved robust sampling technique. Our method periodically uses a proposal distribution constructed from informative range measurements to avoid local maxima. Sampling from this type of proposal distribution leads to new dissemination of samples and avoids wrong modes. 
The key contributions of this paper are as\vspace{-.8mm} follows. 

\begin{itemize}
\item We introduce a proposal distribution for \ac{mva}-based \ac{slam} that can provide robustness in challenging scenarios with range measurements.
\vspace{.8mm}
\item We demonstrate significant performance advantages of the \ac{mva}-based \ac{slam} with robust sampling compared to conventional \ac{mva}-based\vspace{0mm} \ac{slam}.
\end{itemize}

\section{System model of \ac{mva}-based \ac{slam}}
\label{sec:environmentalModel}

We consider a mobile agent and $J$ \acp{pa} with known positions $\V{p}_{\mathrm{pa}}^{(j)} \rmv\!\in\rmv \mathbb{R}^2\rmv$, $j \rmv=\rmv 1,\ldots,J$, where $J$ is assumed to be known. At each discrete time slot $n$, the position of the mobile agent $\V{p}_n \rmv\!\in\rmv \mathbb{R}^2$ is unknown. The mobile agent transmits a radio signal and the \acp{pa} act as receivers. However, the proposed algorithm can be easily reformulated for the case where the \acp{pa} act as transmitters and the mobile agent acts as a receiver.  Associated with the $j$th \ac{pa}, there are $K$ \acp{va} \cite{LeiMeyHlaWitTufWin:J19}  at unknown positions $\V{p}_{k,\mathrm{va}}^{(j)} \!\rmv\in\rmv \mathbb{R}^2\rmv$, $k\rmv=\rmv 1,\ldots,K\!$. \ac{va} position $\V{p}_{k,\mathrm{va}}^{(j)}$ is the mirror image of $\V{p}_{\mathrm{pa}}^{(j)}$ at reflective surface $k$; it represents the single-bounce propagation path agent -- surface $k$ -- \ac{pa} $j$. In a preprocessing stage, distance measurements $d_{m,n}^{(j)}$, $m = 1,\dots,M^{(j)}_n$ are extracted from the radio signal received at \ac{pa} $j$ and time $n$ \cite{LeiMeyHlaWitTufWin:J19,LeitingerICC2019,BadiuTSP2017}.  The distance measurement related to the propagation path represented by $\V{p}_{k,\mathrm{va}}^{(j)}$ is modeled \vspace{0mm} as $z_{m,n}^{(j)} = \big\|  \V{p}_n - \V{p}^{(j)}_{k,\mathrm{va}} \big\| + \nu_{m,n}^{(j)}$ where $\|\cdot\|$ represents the Euclidean norm and $\nu_{m,n}^{(j)}$ is zero-mean Gaussian measurement noise with standard deviation $\sigma^{(j)}_{m,n}$. Note that $\nu_{m,n}^{(j)}$ is assumed statistically independent of $\V{p}_n$ and $\V{p}^{(j)}_{k,\mathrm{va}}$.

A reflective surface is involved in multiple propagation paths and thus gives rise to multiple \acp{va}. To enable the consistent combination, i.e., ``fusion'' of map information provided by distance measurement of different \acp{pa}, following \cite{LeiMey:Asilomar2020_DataFusion}, we represent reflective surfaces by $K$ unique \acp{mva} at unknown positions $\V{p}_{k,\mathrm{mva}} \!\rmv\in\rmv \mathbb{R}^2\rmv$, $k\rmv=\rmv 1,\ldots,K\!$.  The unique \ac{mva} position $\V{p}_{k,\mathrm{mva}} \!\rmv\in\rmv \mathbb{R}^2\rmv$ is defined as the mirror image of the global origin $[0 \hspace{.6mm} 0]^{\T}$ on the reflective surface \vspace{-9mm}$K$.
\begin{figure}[h!]
\centering
\includegraphics[scale=.62]{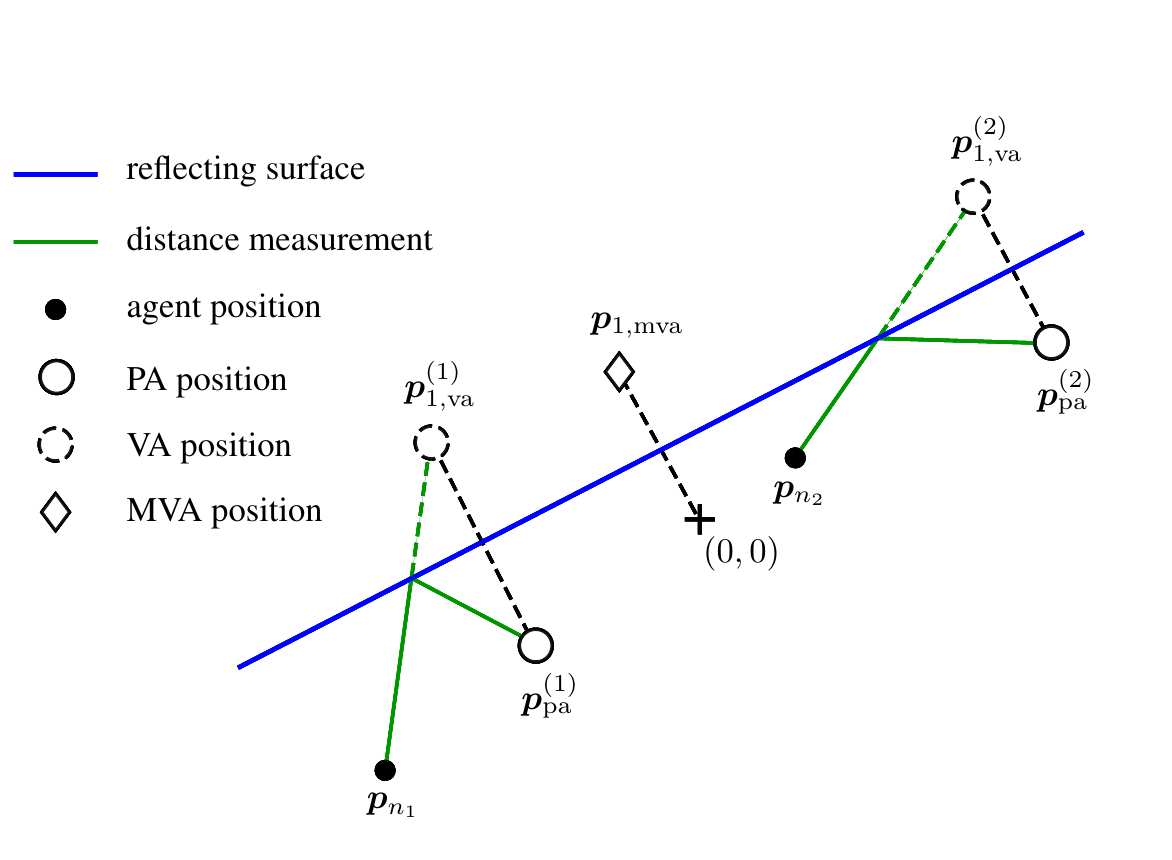}
\vspace{-4mm}
\caption{A graphical depiction of a multipath-based localization scenario involving $K \rmv=\rmv 1$ reflective surfaces, $J \rmv=\rmv 2$ \acp{pa}, and positions of the agent $\V{p}_n$ at two time steps $n \rmv\in\rmv \{n_1, n_2\}$.}
\label{fig:bounce}
\vspace{-1.5mm}
\end{figure}

By using some algebra, the transformation from \ac{mva} $\V{p}_{k,\mathrm{mva}}$ to \ac{va} position $\V{p}^{(j)}_{\mathrm{pa}}$ can be obtained as
\begin{align}
\V{p}^{(j)}_{k,\mathrm{va}} &= -\bigg(\frac{ 2 \big\langle \V{p}_{k,\mathrm{mva}},\V{p}^{(j)}_{\mathrm{pa}} \big\rangle}{ \big\|\V{p}_{k,\mathrm{mva}}\big\|^2 } - 1 \bigg) \ist \V{p}_{k,\mathrm{mva}} + \ist \V{p}^{(j)}_{\mathrm{pa}} \nn\\[1.5mm]
&= h\big( \V{p}_{k,\mathrm{mva}}, \V{p}^{(j)}_{\mathrm{pa}} \big). \label{eq:nonLinearTransformation}
\end{align}
where $\langle \cdot, \cdot \rangle$ represents the inner-product between two vectors.

Based on the nonlinear transformation in \eqref{eq:nonLinearTransformation}, the distance measurement related to the propagation path agent -- surface $k$ -- \ac{pa} $j$ can now alternatively be expressed \vspace{.5mm} by \cite{LeiMey:Asilomar2020_DataFusion}
\begin{align}
z_{m,n}^{(j)}&= \big\|  \V{p}_n - h\big( \V{p}_{k,\mathrm{mva}}, \V{p}^{(j)}_{\mathrm{pa}} \big) \big\| + \nu_{m,n}^{(j)}. \label{eqB} \\[-5mm]
\nn
\end{align}
Using \eqref{eqB} as well as the known \ac{pa} position $\V{p}^{(j)}_{\mathrm{pa}}$, we can directly obtain the \ac{mva}-based likelihood function as $f\big(z_{m,n}^{(j)}\big|\V{p}_{k,\mathrm{mva}},\V{p}_{n}\big)\vspace{.8mm}$. Note \vspace{-.5mm} that with the proposed \ac{mva}-based measurement model, at time $n$, the measurements collected by all \acp {pa} $j \rmv=\rmv 1,\ldots,J$ can provide information on the same \ac{mva} and agent positions $\V{p}_{k,\mathrm{mva}}, k \in \{1,\dots,K\}$ and $\V{p}_{n}$, respectively. The number of \acp{mva} $K$ is unknown. As an example, Fig.~\ref{fig:bounce} shows a scenario with two \acp{pa} $j \in \{1,2\}$ and two time steps $n_1$ and $n_2$. Note that due to measurement origin uncertainty there is an unknown association of measurements with \acp{mva}. Furthermore there can be missed detections, i.e., at some time steps $n$ actual \acp{mva} may not produce a measurement at some \ac{pa} $j$; and false alarms, i.e., there may be clutter measurements not generated by any \ac{mva} (see \cite{LeiMeyHlaWitTufWin:J19,LeitingerICC2019,MenMeyBauWin:J19,MeyWil:J21} for details).

At each time $n$, the state of the agent $\V{x}_{n}$ consists of it's position $\V{p}_{n}$ and possibly further motion related parameters.  As in \cite{MeyBraWilHla:J17,MeyKroWilLauHlaBraWin:J18,LeiMeyHlaWitTufWin:J19}, we account for the unknown number of \acp{mva} by introducing \acp{pmva} $k \in \{ 1,\dots, K_n \}$. The number $K_n$ of \acp{pmva} is the maximum possible number of actual \ac{mva} that produced a measurement so far \cite{MeyKroWilLauHlaBraWin:J18} (where $K_n$ increases with time). \ac{pmva} states are denoted as $\V{y}_{k,n} \rmv=\rmv \big[\V{p}^{\T}_{k,\mathrm{mva}} \; r_{k,n} \big ]^\T\rmv\rmv\rmv$.
The existence/nonexistence of \ac{pmva} $k$ is modeled by the existence variable $r_{k,n} \rmv\in \{0,1\}$ in the sense that \ac{pmva} $k$ exists if and only if $r_{k,n} \!=\! 1$. It is considered formally also if \ac{pmva} $k$ is nonexistent, i.e., if $r_{k,n} \!=\rmv 0$. The states $\V{p}^{\T}_{k,\mathrm{mva}}$ of nonexistent \acp{pmva} are obviously irrelevant. Therefore, all \acp{pdf} defined for \ac{pmva} states, $f(\V{y}_{k,n}) =\rmv f(\V{p}_{k,\mathrm{mva}}, r_{k,n})$, are of the form $f(\V{p}_{k,\mathrm{mva}}, 0 )$ $=\rmv f_{k,n} f_{\text{d}}(\V{p}_{k,\mathrm{mva}})$, where $f_{\text{d}}(\V{p}_{k,\mathrm{mva}})$ is an arbitrary ``dummy \ac{pdf}'' and $f_{k,n} \!\rmv\in [0,1]$ is a constant. Further details for the system model of \ac{mva}-based RF-SLAM are provided in \cite{LeiMey:Asilomar2020_DataFusion}.

\section{Problem Formulation and Proposed Method}
\label{sec:probMethod}

We aim to estimate the agent state $\V{x}_n$ using all available measurements $\V{z}_{1:n}\rmv$ from all \acp{pa} up to time $n$. In particular, we calculate an estimate $\hat{\V{x}}_n$ by using the \ac{mmse} estimator \vspace{.5mm} \cite[Ch.~4]{Poo:B94}
\begin{equation}
\hat{\V{x}}_n \ist\triangleq\, \int \rmv \V{x}_n \ist f( \V{x}_n|\V{z}_{1:n}) \ist \mathrm{d} \V{x}_n \rmv. \label{eq:mmse1}
\vspace{.5mm}
\end{equation}
For the mapping of reflection surfaces, detection of \acp{pmva} $k \!\in\! \{ 1,\dots,K_n \}$ and estimation their positions $\V{p}_{k,\mathrm{mva}}$ is considered. This relies on the marginal posterior existence probabilities $p(r_{k,n} \!=\! 1|\V{z}_{1:n})$ and the marginal posterior \acp{pdf} $f(\V{x}_{k,n} | r_{k,n} \!=\! 1, \V{z}_{1:n} )$. A \ac{pmva} $k$ is declared to exist if $p(r_{k,n} \!=\! 1|\V{z}_{1:n}) > p_{\text{de}}$, where $p_{\text{de}}$ is a detection threshold \cite[Ch.~2]{Poo:B94}. The number $\hat{K}_n$ of \ac{pmva} states that are considered to exist is the estimate of total number $K$ of \acp{mva}. For existing \acp{pmva}, an estimate of it's position $\V{p}_{k,\mathrm{mva}}$ can again be calculated by the \ac{mmse} \cite[Ch.~4]{Poo:B94}
\vspace{1mm}
\begin{equation}
\hat{\V{p}}_{k,\mathrm{mva}} \,\triangleq \int \rmv \V{p}_{k,\mathrm{mva}}  \ist\ist f(\V{p}_{k,\mathrm{mva}} \ist | \ist r_{k,n} \!=\! 1, \V{z}_{1:n}) \ist\ist \mathrm{d} \V{p}_{k,\mathrm{mva}} \rmv. \label{eq:mmse2}
\vspace{.5mm}
\end{equation}
The calculation of $f( \V{x}_n|\V{z}_{1:n})$, $p(r_{k,n} \!=\! 1 |\V{z})$, and $f(\V{p}_{k,\mathrm{mva}} |$ $r_{k,n} \rmv=\rmv 1, \V{z}_{1:n})$ from the joint posterior \ac{pdf} \cite[Eq.~(4)]{LeiMey:Asilomar2020_DataFusion} by direct marginalization is not feasible. 

By performing sequential sample-based message passing by means of the \ac{spa} rules \cite{KscFreLoe:01,MeyHliHla:J16,MeyBraWilHla:J17,LeiMeyHlaWitTufWin:J19} on the factor graph in \cite[Fig.~3]{LeiMey:Asilomar2020_DataFusion}, approximations (``beliefs'') of the marginal posterior \acp{pdf} $q( \V{x}_n) \approx f( \V{x}_n|\V{z}_{1:n})$ and $q( \V{p}_{k,\mathrm{mva}}, r_{k,n}) \approx f(\V{p}_{k,\mathrm{mva}}, r_{k,n}  | \V{z}_{1:n})$, $k \in \{1,\dots,K_n\}$  can be obtained in an efficient ways. More specifically, representations of beliefs that consist of $I$ weighted random samples or particles denoted as $\big\{ \V{x}^{(i)}_n\rmv\rmv, \mathpzc{w}_n^{(i)} \big\}^{I}_{i=1} \sim q( \V{x}_n)$ and $\big\{ \big( \V{p}^{(i)}_{k,\mathrm{mva}}, w_{k,n}^{(i)} \big) \big\}^{I}_{i=1} \sim q( \V{p}_{k,\mathrm{mva}}, r_{k,n})$, $k \in \{1,\dots,K_n\}$ are computed \cite{MeyHliHla:J16,MeyBraWilHla:J17,LeiMeyHlaWitTufWin:J19}. Note that $\sum^{I}_{i=1}\mathpzc{w}_{n}^{(i)} = 1$ while $\sum^{I}_{i=1}w_{k,n}^{(i)} \approx p(r_{k,n} \!=\! 1 |\V{z})$ (cf. \cite[Sec.~VI]{MeyBraWilHla:J17}). These sample-based representations can be used for approximate MMSE estimation by evaluating \eqref{eq:mmse1} and \eqref{eq:mmse2} based on Monte Carlo integration \cite{doucet01}.  To avoid the number of \ac{pmva} states growing indefinitely, \acp{pmva} states with $p( r_{k,n} \rmv= 1  \ist | \ist \V{z}_{1:n})$ below a threshold $p_{\text{pr}}$ are removed from the state space (``pruned''). Pruning is performed at each time $n$, after the measurements of all \acp{pa} have been processed.
\vspace{1mm}

\section{Review of Bootstrap Sampling for \ac{slam}}
\label{sec:reviewBootstrap}

Existing \ac{slam} methods employ the bootstrap sampling strategy \cite{AruMasGorCla:02} where predicted beliefs are employed as proposal \ac{pdf}.  In particular, the proposal distribution for calculating the belief $q( \V{p}_{k,\mathrm{mva}}, r_{k,n})$ of \ac{mva}, $k \in \{1,\dots,K_n\}$ at time $n$ is given by (cf.~\cite{MeyHliHla:J16,LeiMeyHlaWitTufWin:J19,LeiMey:Asilomar2020_DataFusion})
\begin{equation}
f_{\mathrm{pred.}}(\V{x}_{n}, \V{p}_{k,\mathrm{mva}}) \propto \alpha( \V{p}_{k,\mathrm{mva}}, r_{k,n} \rmv=\rmv 1) \ist \alpha( \V{x}_{n}) \label{eq:proposal}
\end{equation}
where $\propto$ indicates equality up to a normalization factor and the ``prediction messages'' $\alpha( \V{p}_{k,\mathrm{mva}}, r_{k,n})$ and $\alpha( \V{x}_{n})$ can be obtained as
\begin{align}
 \alpha( \V{x}_n) &= \int \rmv f(\V{x}_n|\V{x}_{n-1}) \ist q(\V{x}_{n-1}) \ist \mathrm{d}\V{x}_{n-1} \nn\\
  \alpha( \V{p}_{k,\mathrm{mva}},\rmv r_{k,n}) &=\rrmv\rrmv\rrmv  \sum_{r_{k,n-1} \in \{0,1\}}\int \rmv\rmv f\big(\V{p}_{k,\mathrm{mva}}\ist, r_{k,n} \ist \big| \ist \V{p}_{k,\mathrm{mva}}\ist, r_{k,n-1} \big) \nn\\[0mm]
 &\hspace{15mm}\times q\big( \V{p}_{k,\mathrm{mva}}, r_{k,n-1} \big) \ist \mathrm{d} \V{p}_{k,\mathrm{mva}} \ist.
 \label{eq:stateTransitionMessageFeature} 
\vspace{-1.2mm}
\end{align}
Here, $f(\V{x}_n|\V{x}_{n-1})$ and $f\big(\V{p}_{k,\mathrm{mva}}\ist, r_{k,n} \ist \big| \ist \V{p}_{k,\mathrm{mva}}\ist, r_{k,n-1} \big)$ are the state-transition functions for the agent state and the \ac{mva} state, respectively \cite[Sec.~\ref{sec:environmentalModel}]{LeiMey:Asilomar2020_DataFusion}. Furthermore, the beliefs of the agent state, $q(\V{x}_{n-1})$, and of the \ac{mva} states $q\big( \V{p}_{k,\mathrm{mva}}, r_{k,n-1} \big)$, were calculated at the preceding time $n \rmv-\! 1$. Note that for the proposal $f_{\mathrm{pred.}}(\V{x}_{n}, \V{p}_{k,\mathrm{mva}})$ in \eqref{eq:proposal}, we only use the functional form of $\alpha( \V{p}_{k,\mathrm{mva}}, r_{k,n})$ for the case $r_{k,n} \rmv=\rmv 1$, since  the functional form for the case $r_{k,n} \rmv=\rmv 0$ is always equal to the dummy \ac{pdf} $f_{\text{d}}(\V{p}_{k,\mathrm{mva}})$. Samples of $f_{\mathrm{pred.}}( \V{x}_{n}, \V{p}_{k,\mathrm{mva}})$ can be obtained as discussed in \cite{MeyBraWilHla:J17,LeiMeyHlaWitTufWin:J19}. 

The proposal distribution in \eqref{eq:proposal} has to be a function of both \ac{mva} position $\V{p}_{k,\mathrm{mva}}$ and  agent state $\V{x}_n$, since the measurement model \eqref{eqB} also involves both \ac{mva} state and agent state. Furthermore, note that the proposal distribution in \eqref{eq:proposal} is also used to calculate a factor that represents the contribution of \ac{mva}  $k \in \{1,\dots,K_n\}$ to the agent weights $\mathpzc{w}_n^{(i)}\rmv\rmv$, $i \rmv\in\rmv \{1,\dots,I\}$ (cf.~\cite{MeyHliHla:J16,LeiMeyHlaWitTufWin:J19,LeiMey:Asilomar2020_DataFusion}). 

Bootstrap sampling is suitable for \ac{slam} methods that consider \acp{va} as the features to be mapped \cite{LeiMeyHlaWitTufWin:J19,LeitingerICC2019,MenMeyBauWin:J19}. However, its use for the fast and more accurate \acp{mva}-based \ac{slam} is problematic in challenging \ac{slam} scenarios with one or two \acp{pa} where only range measurements are available. This is because, due to certain geometric symmetries, for some initial time steps after an \ac{mva} has been introduced, the \acp{pdf} of an \ac{mva} can be multi-modal. In addition, during these initial time steps, often, the dominant mode might not be the one located at the correct \ac{mva} position. Thus, the sample representations provided by bootstrap sampling may collapse in a wrong mode, i.e., converge to incorrect \ac{mva} positions corresponding to local maxima.
\vspace{1mm}

\section{Robust Sampling for \ac{mva}-based \ac{slam}}
\label{sec:robustSampling}

To address these limitations, we introduce an alternative strategy for obtaining a proposal distribution. In what follows, we will again discuss the proposal distribution for the belief $q( \V{p}_{k,\mathrm{mva}}, r_{k,n})$ of \ac{mva}, $k \in \{1,\dots,K_n\}$. As discussed later this proposal will just be used at certain time steps $n$. 

The main idea is to construct a weighted mixture that consists of \ac{mva} position information that is predicted and \ac{mva} position information that is related to the best measurement for each \ac{pa} $j \rmv=\rmv 1,\ldots,J$ obtained at time step $n\rmv-\rmv1$. A measurement is the best if it has the largest probability of association with \ac{mva} $k$ at time $n\rmv-\rmv1$. The index\vspace{.3mm}  of this best measurement is denoted\vspace{-.7mm} by $m^{(j)}_{k,n-1} \in \big\{0,1,\dots,M^{(j)}_{n-1}\big\}$, where $m^{(j)}_{k,n-1} = 0$ indicates that no measurement is the best, i.e., the probability of no measurement being associated to the \ac{mva} (``missed detection'') is larger than the probability of association with any measurement. To construct the proposal distribution at time $n$, we use the measurements from time $n-1$ because association probabilities at time $n-1$ have already been calculated \cite{LeiMeyHlaWitTufWin:J19}; on the other hand, calculation of association probabilities at time $n$ is based on the proposal distribution at time $n$.

The considered proposal distribution at time $n$, is given\vspace{-.5mm} by 
\begin{align}
&f_{\mathrm{mixture}}(\V{x}_{n}, \V{p}_{k,\mathrm{mva}})  \nn\\[1.5mm]
& \hspace{6mm}\triangleq  f_{\mathrm{pred.}}(\V{x}_{n}, \V{p}_{k,\mathrm{mva}}) + \sum_{j\in\Set{J}} f^{(j)}_{\mathrm{meas.}}(\V{x}_{n}, \V{p}_{k,\mathrm{mva}}) \label{eq:finalProposal}\\[-8mm]
\nn
\end{align}
where $\Set{J}$ consists of all indexes $j$ with $m^{(j)}_{k,n-1} \rmv\neq 0$. Furthermore, the component related to the best measurements\vspace{1.3mm} reads
\begin{align}
&f^{(j)}_{\mathrm{meas.}}( \V{x}_{n}, \V{p}_{k,\mathrm{mva}}) \propto\rmv \alpha( \V{x}_{n}) f \big(z_{m^{(j)}_{k,n-1},n-1}^{(j)}\big| \V{p}_{k,\mathrm{mva}}, \V{p}_{n}\big). \nn
\end{align}

Samples of $f^{(j)}_{\mathrm{meas.}}( \V{x}_{n}, \V{p}_{k,\mathrm{mva}})$ can be obtained  by following the importance sampling principle \cite{doucet01}. In particular, samples $\big\{( \tilde{\V{x}}^{(i)}_{n}, \tilde{\V{p}}^{(i)}_{k,\mathrm{mva}})\big\}^{I'}_{i=1} $ are drawn from $ \alpha( \V{x}_{n}) \ist f_{\mathrm{uni.}} (\V{p}_{k,\mathrm{mva}})$ first, where $f_{\mathrm{uni.}} (\V{p}_{k,\mathrm{mva}})$ is a \ac{pdf} that is uniform on the area of interest. Then corresponding unnormalized weights are calculated\vspace{-1mm} as 
\begin{equation}
\tilde{w}^{(i)}_{k,n} = \frac{f \Big(z_{m^{(j)}_{k,n-1},n-1}^{(j)} \ist\big| \ist\tilde{\V{p}}^{(i)}_{k,\mathrm{mva}}, \tilde{\V{p}}^{(i)}_{n}\Big)}{f_{\mathrm{uni.}} \big(\tilde{\V{p}}^{(i)}_{k,\mathrm{mva}}\big)}.
\end{equation}
The samples $\big\{( \V{x}^{(i)}_{n}, \V{p}^{(i)}_{k,\mathrm{mva}})\big\}^{I'}_{i=1} $ representing $f^{(j)}_{\mathrm{meas.}}( \V{x}_{n},$ $\V{p}_{k,\mathrm{mva}})$ are computed from $\big\{( \tilde{\V{x}}^{(i)}_{n}, \tilde{\V{p}}^{(i)}_{k,\mathrm{mva}}, \tilde{w}^{(i)}_{k,n})\big\}^{I'}_{i=1} $ by first normalizing the weights $\tilde{w}^{(i)}_{k,n}$, $i \in \{1,\dots,I'\}$ and then performing a resampling step \cite{doucet01}. Finally, $I$ samples of the considered proposal distribution $f_{\mathrm{mixture}}(\V{x}_{n}, \V{p}_{k,\mathrm{mva}}, r_{k,n})$ in \eqref{eq:finalProposal} are obtained as follows: (i) $I'$ is selected as small as possible such that $I' (|\Set{J}| + 1) \geq I$; (ii) $I'$ samples from $f_{\mathrm{pred.}}(\V{x}_{n}, \V{p}_{k,\mathrm{mva}})$  and from each $f^{(j)}_{\mathrm{meas.}}(\V{x}_{n}, \V{p}_{k,\mathrm{mva}})$, $j \rmv\in\rmv \Set{J}$, are obtained; and (iii) $I$ of the resulting $I' (|\Set{J}| + 1)$ samples are selected randomly.

The main advantage of using \eqref{eq:finalProposal} as the proposal distribution is that particles are again spread out over a wider area of potential \ac{mva} locations and thus collapsing to a wrong mode is avoided. On the other hand, using \eqref{eq:finalProposal} at each time step would significantly reduce the speed of convergence to the correct mode. In our numerical evaluation of \ac{mva}-based \ac{slam}, we found it useful to only use the proposal distribution \eqref{eq:finalProposal} at certain time steps $n'$ that are determined randomly. Let $n_1$ be the last time steps where \eqref{eq:finalProposal} has been used. The next time step $n_2$ where \eqref{eq:finalProposal} is used can now be obtained by sampling from the uniform \ac{pmf} $p_{\mathrm{uni.}}\big(n_2;n_1+N_{\mathrm{1}},n_1+N_{\mathrm{2}}\big)$. Finally, the proposal \eqref{eq:finalProposal} is not used anymore after the \ac{pmva} has existed for a certain number of time steps $N_{\mathrm{max}}$.  A possible choice for the hyperparameters $N_{\mathrm{1}}$, $N_{\mathrm{2}}$ are $N_{\mathrm{max}}$ is discussed in Section\vspace{0mm}~\ref{sec:results}. When we calculate the weights of the samples drawn from the proposal \eqref{eq:finalProposal}, we perform an approximation and avoid the costly evaluation of \eqref{eq:finalProposal}. As demonstrated in Section \ref{sec:results}, despite this approximation the proposed robust sampling can yield convincing \ac{mva}-based \ac{slam} performance.

\section{Simulation Results}
\label{sec:results}

 \begin{figure}[t!]
 \centering
 \includegraphics[width=0.73\columnwidth]{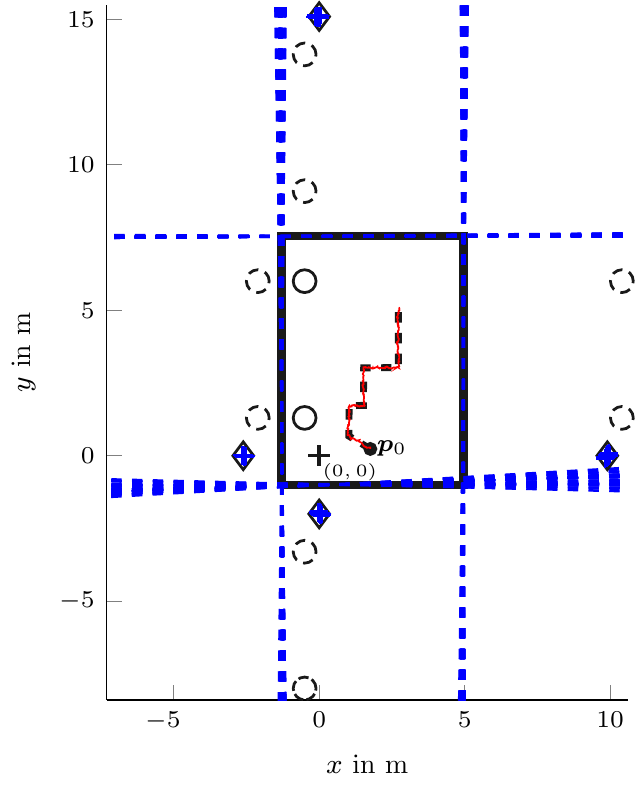}\\[2mm]
 \centering
 \includegraphics[width=0.73\columnwidth]{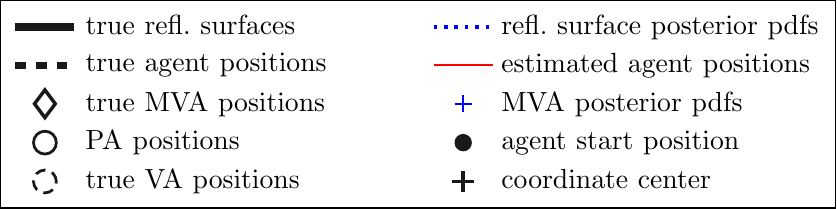}
 \caption{Considered scenario for performance evaluation with three \acp{pa}, four reflective surfaces and corresponding \acp{mva} and \acp{va}, as well as agent trajectory.}\label{fig:floorplan}
 \vspace{-4mm}
 \end{figure}

The proposed robust sampling for \ac{mva}-based \ac{slam} is validated in the indoor scenario shown in Figure~\ref{fig:floorplan}. The scenario consists of four reflective surfaces, i.e., $K=4$ \acp{mva}, as well as two \acp{pa} at positions $\V{p}_{\mathrm{pa}}^{(1)} = [-0.5\iist\iist 6]^{\T}$ and $\V{p}_{\mathrm{pa}}^{(3)} = [-0.5\iist\iist 1.3]^{\T}$. We compare the proposed method with \ac{mva}-based \ac{slam} that relies on bootstrap sampling \cite{LeiMey:Asilomar2020_DataFusion}.

The agent's state-transition \ac{pdf} $f(\V{x}_n|\V{x}_{n-1})$, with $\V{x}_n \rmv=\rmv  [\V{p}_n^{\T} \; \V{v}_n^{\T} ]^{\T}\rmv$, is defined by a linear, near constant-velocity motion model \cite[Sec. 6.3.2]{BarShalom2002EstimationTracking}, i.e., $\V{x}_n =  \V{A} \V{x}_{n-1} + \V{B} \V{\omega}_n$.
Here, $\V{A} \rmv\in\rmv \mathbb{R}^{4 \times 4}$ and $\V{B} \rmv\in\rmv \mathbb{R}^{4 \times 2}$ are as defined in \cite[Sec. 6.3.2]{BarShalom2002EstimationTracking} 
(with sampling period $\vu{\Delta T \rmv=\rmv 1\rmv}{s}$), and the driving process $\V{\omega}_n$ is iid across $n$, zero-mean, and Gaussian with covariance matrix $\sigma_{\omega}^2\bold{I}_2$, where $\bold{I}_2$ denotes the $2 \rmv\times\rmv 2$ identity matrix and $\sigma_{\omega} = 0.0032\,\text{m}/\text{s}^2$. For the sake of numerical stability, we introduced a small regularization noise to the \ac{pmva} state $\V{p}_{k,\mathrm{mva}}$ at each time $n$, i.e., $\underline{\V{p}}_{k,\mathrm{mva}} \rmv\rmv=\rmv\rmv \V{p}_{k,\mathrm{mva}} \rmv+\rmv \V{a}_{k}$, where $\V{a}_{k}$ is iid across $k$, zero-mean, and Gaussian with covariance matrix $\sigma_{\V{a}}^2\bold{I}_2$ and $\sigma_{\V{a}} = 10^{-5}\,\text{m}$. 

We performed 300 simulation runs using 30,000 samples, each using the floor plan and agent trajectory shown in Fig.~\ref{fig:floorplan}. In each simulation run, we generated noisy distance measurements $z_{m,n}^{(j)}$ according to \eqref{eqB} with noise standard deviation $\sigma_{m,n}^{(j)} = 0.1\,$m and detection probability $p_\mathrm{d} = 0.95$. In addition, a mean number $\mu_{\mathrm{fa}} = 1$ of false alarm measurements $z_{m,n}^{(j)}$ were generated according to a false alarm \ac{pdf} $f_{\mathrm{fa}}(z_{m,n}^{(j)})$ that is uniform on $[0\,\text{m} \ist\ist\ist 30\,\text{m}]$. The samples for the initial agent state are drawn from a 4-D uniform distribution with center $\V{x}_0 = [\V{p}_{0}^{\T}\;0\;\, 0]^{\T}\rmv$, where $\V{p}_{0}$ is the starting position of the actual agent trajectory. The support of each position component about the respective center is given by $[-0.1\,\text{m}, 0.1\,\text{m}]$ and of each velocity component is given by $[-0.01\,\text{m/s}, 0.01\,\text{m/s}]$. At time $n \rmv=\rmv 0$, the number of \acp{mva} is $K_0 = 0$, i.e., no prior map information is available. The prior distribution for new \ac{pmva} states $f_\mathrm{n}(\overline{\V{y}}_{m,n})$ is uniform on the square region given by $[-15\,\text{m},\ist 15\,\text{m}]\ist\times\ist[-15\,\text{m},\ist 15\,\text{m}]$ around the center of the floor plan shown in Fig.~\ref{fig:floorplan} and the mean number of new \ac{pmva} at time $n$ is $\mu_n = 0.01$. The probability of survival is $p_{\mathrm{s}} = 0.999$, the detection threshold is $p_{\mathrm{de}} = 0.5$, and the pruning threshold is $p_{\mathrm{pr}} = 10^{-3}$. The parameters for robust sampling are $N_1 = 5$, $N_2 = 10$, and $N_\text{max} = 120$, respectively.

 \begin{figure}[h!]
 \centering
 \subfloat[\label{fig:MVAserror}]{\includegraphics[width=1\columnwidth, height=0.38\columnwidth]{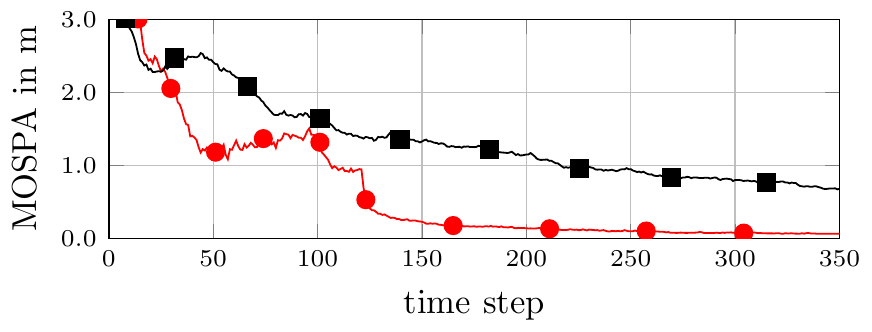}}\\[-1mm]
 \subfloat[\label{fig:agenterror}]{\includegraphics[width=1\columnwidth,height=0.38\columnwidth]{Fig3_a}}\\[1mm]
 \centering
 \includegraphics[width=0.73\columnwidth]{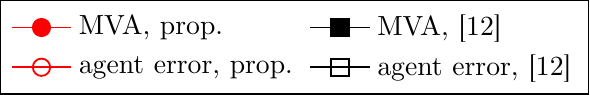}
 \caption{Preliminary performance results: (a) \ac{rmse} position errors of the \acp{mva} and (b) \ac{rmse} of the mobile agent.
 \vspace{-10mm}
 }
 \end{figure}

As an example, Fig.~\ref{fig:floorplan} depicts for one simulation run the posterior \acp{pdf} represented by samples of the \ac{mva} positions and corresponding reflective surfaces as well as estimated agent tracks. Fig.~\ref{fig:MVAserror} shows the \ac{mospa} errors \cite{Schuhmacher2008} of the \ac{mva} positions, all versus time $n$. The \ac{mospa} errors are based on the Euclidean metric with cutoff parameter $c = 5\,$m and order $p = 1$. The red line shows the \ac{mospa} errors of the proposed \ac{mva}-based SLAM algorithm and the black line shows the \ac{mospa} errors of the algorithm presented in \cite{LeiMey:Asilomar2020_DataFusion}. The \ac{mospa} error of the proposed \ac{mva}-based SLAM algorithm converges to a much smaller mapping error than that of the algorithm presented in \cite{LeiMey:Asilomar2020_DataFusion}. This can be explained by the symmetric multi-modality of the marginal posterior \acp{pdf}. The proposed robust sampling avoids the behavior where all samples collapse into the wrong mode. However, since the robust sampling ``excites'' the multi-modality of the marginal \ac{pdf} of the \acp{mva}, the \ac{mospa} error remains quite large until robust sampling is disabled at time $n = 120$. Finally, Fig.~\ref{fig:agenterror} shows the \acp{rmse} of the agent positions of the converged simulation runs versus time $n$. We define a simulation run to be converged if $\{\forall n: ||\hat{\V{x}}_n - \V{x}_n|| < 0.5 \}$. For the converged runs both methods have an agent \ac{rmse} below $0.1\,$m, however, the agent \ac{rmse} provided by the proposed algorithms is still significantly smaller. More importantly, only one of the $300$ simulation runs diverged for the proposed algorithm, but $38\,$\% of the simulation runs diverged for the algorithm in \cite{LeiMey:Asilomar2020_DataFusion}.

\section{Conclusions and Future Work}

In this paper, we introduced \ac{mva}-based \ac{slam} with an improved sampling technique that is suitable for challenging scenarios where only range measurements are available, and only one or two \acp{pa} are deployed. Our numerical evaluation demonstrated significant performance advantages of the proposed method compared to the recently introduced conventional \ac{mva}-based \ac{slam}. Promising directions for future research are an extension of \ac{mva}-based \ac{slam} to (i) angle measurements provided by antenna arrays and (ii) higher-order reflections from flat\vspace{.5mm} surfaces.


\renewcommand{\baselinestretch}{0.97}
\selectfont
\section*{Acknowledgement}
{ \small
DISTRIBUTION STATEMENT A: Approved for public release. This work was supported in part by the Under Secretary of Defense for Research and Engineering under Air Force Contract No. FA8702-15-D-0001. Any opinions, findings, conclusions, or recommendations expressed in this material are those of the author(s) and do not necessarily reflect the views of the Under Secretary of Defense for Research and\vspace{.5mm} Engineering. This work was also supported in part by the Christian Doppler Research Association, the Austrian Federal Ministry for Digital and Economic Affairs and the National Foundation for Research, Technology and Development. }

\bibliographystyle{IEEEtran}
\bibliography{IEEEabrv,StringDefinitions,Books,Papers,References}

\end{document}

%% file: pgfplot.tex
\usepackage{ifluatex}

\ifluatex 
    \usepackage{fontspec}
    \defaultfontfeatures{Ligatures={TeX}}

\else
    \usepackage[latin1]{inputenc}
    \usepackage[T2A,T1]{fontenc}
\fi
\usetikzlibrary{backgrounds}
\usetikzlibrary{calc,arrows.meta}
\ifthenelse{\equal{\exportFigures}{true}}
{
  \usepgfplotslibrary{external}\tikzexternalize[prefix=\tikzfolder]
  \tikzset{external/system call={lualatex \tikzexternalcheckshellescape -halt-on-error -interaction=batchmode -jobname "\image" "\texsource"}}
}
{}

\def\addlegendimage{\csname pgfplots@addlegendimage\endcsname}
\def\addlegendentry{\csname pgfplots@addlegendentry\endcsname}

\newcommand{\pgfref}[1]
{\ifthenelse{\equal{\exportFigures}{true}}
{\tikzexternaldisable\ref{#1}\tikzexternalenable}
{\ref{#1}}}

\pgfplotsset{compat=newest} 

\newlength\figureheight 
\newlength\figurewidth 
\usetikzlibrary{patterns,arrows}
\usepgfplotslibrary{colormaps}
\usetikzlibrary{plotmarks}
\pgfplotsset{every axis/.append style={
  label style={font=\normalsize},
  tick label style={font=\scriptsize},
  xticklabel={
   \ifdim \tick pt < 0pt
    \pgfmathparse{abs(\tick)}%
    \llap{$-{}$}\pgfmathprintnumber{\pgfmathresult}
   \else
    \pgfmathprintnumber{\tick}
   \fi}
   }}
\usetikzlibrary{decorations.markings}
\makeatletter
\tikzset{
  nomorepostactions/.code={\let\tikz@postactions=\pgfutil@empty},
  mymarkfixednumber/.style n args={3}{decoration={markings,
    mark= between positions 0.1 and 1 step (1/#3)*\pgfdecoratedpathlength with{%
        \tikzset{#2,every mark}\tikz@options
        \pgfuseplotmark{#1}%
      },  
    },
    postaction={decorate},
    /pgfplots/legend image post style={
        mark=#1,mark options={#2},every path/.append style={nomorepostactions}
    },
  },
}
\tikzset{
  nomorepostactions/.code={\let\tikz@postactions=\pgfutil@empty},
  mymarkfixeddistance/.style n args={3}{decoration={markings,
    mark= between positions 0 and 1 step #3cm with{%
        \tikzset{#2,every mark}\tikz@options
        \pgftransformresetnontranslations%
        \pgfuseplotmark{#1}%
      },  
    },
    postaction={decorate},
    /pgfplots/legend image post style={
        mark=#1,mark options={#2},every path/.append style={nomorepostactions}
    },
  },
}
\tikzset{
  nomorepostactions/.code={\let\tikz@postactions=\pgfutil@empty},
  mymark/.style n args={3}{decoration={markings,
    mark= between positions 0 and 1 step 0.75cm with{%
        \tikzset{#2,every mark}\tikz@options
        \pgftransformresetnontranslations%
        \pgfuseplotmark{#1}%
      },  
    },
    postaction={decorate},
    /pgfplots/legend image post style={
        mark=#1,mark options={#2},every path/.append style={nomorepostactions}
    },
  },
}
\makeatother

\definecolor{colA}{rgb}{0,0,1}
\definecolor{colB}{rgb}{1.00000,0.0,0.00000}
\definecolor{colC}{rgb}{0.1333,0.5451,0.1333}
\definecolor{colD}{rgb}{0.9,0.0,0.9}
\definecolor{colE}{rgb}{0,1,1}
\definecolor{colF}{rgb}{1.00000,0.55,0.00000}

\def\mymarksize{1.5}
\def\mymarksize2{2.5}

\pgfdeclareplotmark{markarrow}
{%
  \pgfpathmoveto{\pgfqpoint{\pgfplotmarksize}{0pt}}
  \pgfpathlineto{\pgfqpointpolar{130}{1.5\pgfplotmarksize}}
  \pgfpathmoveto{\pgfqpoint{\pgfplotmarksize}{0pt}}
  \pgfpathlineto{\pgfqpointpolar{-130}{1.5\pgfplotmarksize}}
  \pgfusepathqstroke
}